\documentclass{article}

\usepackage{PRIMEarxiv}

\usepackage[utf8]{inputenc} 
\usepackage[T1]{fontenc}    
\usepackage{hyperref}       
\usepackage{url}            
\usepackage{booktabs}       
\usepackage{amsfonts}       
\usepackage{nicefrac}       
\usepackage{microtype} 
\usepackage{amsmath}
\usepackage{float}
\usepackage{fancyhdr}       
\usepackage{graphicx}       
\graphicspath{{media/}} 

\title{Modelling Prepayment and Default under Changing Credit Market Conditions
for a Net Present Value Analysis
}

\author{
 Quirini Lorenzo\\
  Banca Monte dei Paschi di Siena\thanks{This article represents the views of the author and not those of the company}\\
  Siena (Italy) \\
  \texttt{lorenzo.quirini@mps.it}\\
   \And
  Vannucci Luigi\\
  University of Florence \\
  Florence (Italy) \\
  \texttt{luigi.vannucci@unifi.it} \\
  \And
  Quirini Giovanni\\
  University of Florence\thanks{Master student in Statistics and Data Science}\\
  Florence (Italy)\\
  \texttt{giovanni.quirini@edu.unifi.it}   \\
}

\begin{document}
\maketitle

\begin{abstract}
A model is developed to assess the profitability of loans or mortgages structured according to a specified repayment schedule. Financial institutions typically face two competing risks—default and prepayment—both of which are influenced by the stochastic evolution of credit market conditions. This study concentrates on the analysis of the Random Net Present Value (RNPV) as a key performance metric. Specifically, it evaluates the mean and variance of the RNPV at both the individual loan level and the portfolio level, within a unified framework that simultaneously accounts for borrower behavior and prevailing credit market dynamics.
\end{abstract}

\section{Introduction}
Firms in the financial sector commonly rely on models to evaluate borrowers’ creditworthiness and to estimate loan profitability, incorporating both macro-level credit market conditions and individual-level applicant characteristics.

The environment in which a company operates may assume different states, ranging from favorable to unfavorable, systematically influencing applicants' behaviour by altering their default and prepayment intensities, as well as the recovery rate in the event of default.

In this conceptual frame we propose a model, designed for the consumer finance industry, able to measure the profitability for loans with given amortization plans. The profitability of a single loan or a loan portfolio can be measured by the random net present value (RNPV). The uncertainty of the economic result is essentially related to two risks, default and prepayment, and the intensities of such risks depend over time on the changing credit environment. In fact, the dynamics of the conditions in which the lender acts is stochastic, due to macroeconomic factors, as well as the commercial and credit policies carried out by the lender. Here we assume that the changing credit environment evolves according to a Markov Chain.
\\
The ideas that underpin this model stem from the professional experience of some of the authors in developing decision processes that support the lender to evaluate the creditworthiness of its customers. Our approach requires the company to estimate the intensities of default and prepayment conditioned on the evolution of the credit market.
\\
It is important to note that the credit environment encompasses not only macroeconomic conditions but also the lender's commercial and credit policies. In this regard, we refer to Quirini and Vannucci (2011) \cite{quirini2011monitoring}, which presents an application of Hidden Markov Models to describe how a risk indicator evolves over time due to a macroeconomic event and the impact of company-level decisions (the variability of the rejection rate).
\\
We report some references we have found particularly useful about two important topics: competing risks and Markov Chains regarded as a tool able to describe the evolution of the credit market conditions.
\\
For the reader interested in statistical estimation of competing risks as default and prepayment, we refer to classical books like Crowder \cite{crowder2001} \cite{crowder2012}, or a more recent one Bayersman, Schumacher and Allignol \cite{BSA2012}, as suited for R users.
\\
In the literature on the consumer credit industry, survival analysis is generally applied to models with a single cause of failure, represented by the event of default. One of the first paper on this topic was written by Narain (1992) \cite{narain1992survival}. Another influential study from the late 1990s is Basanik, Crook and Thomas (1999) , in which the authors introduced a competing risks approach to loan modeling. Following this field of research and taking under analysis the event of default, we mention Stepanova and Thomas (2002) \cite{stepanova2002survival}, who developed an application score for personal loans. For the use of survival analysis in credit scoring models that incorporates macroeconomic variables, we refer to Bellotti and Crook (2009) \cite{bellotti2009credit}.
\\
The first application of Markov Chains in the specialized literature on consumer credit can be traced back in the pioneering work of Cyert, Davidson and Thompson (1962) \cite{cyert1962estimation}. Since that publication, Markov Chains have been widely applied to model customer behaviour over time and, to some extent, to describe the evolution of credit market conditions, which are often considered hidden.
With respect to this field of application, we refer to the paper written by Crowder, Davis and Giampieri (2005) \cite{crowder2005}, while Zucchini and MacDonald (2009) \cite{zucchini2009hidden} provide a general introduction to Hidden Markov Models.
In a previous work, Quirini and Vannucci \cite{quirini2014hmm}, used a "double Markov Chain" to model both the borrower's creditworthiness, in terms of the payments arrears, and  the evolving credit environment.
\\ 
For more recent contributions to the application of Hidden Markov Models (HMMs) in personal loans, we refer to Firdoos \cite{firdoos}, who proposes a method for predicting loan defaults by combining sentiment analysis with HMMs. Feng-Hui Yu et al. \cite{yu2018hmmcredit} present an intensity-based credit risk model incorporating a Hidden Markov Model alongside a filtering method to infer the hidden states. Additionally, Blümke \cite{bluemke2022shmm} introduces a structured Hidden Markov Model to estimate macroeconomic scenario probabilities for loan loss provisions, reducing dependence on expert judgment.
\\
At the best of our knowledge, our new contribution focuses on  measuring actual profitability - in terms of the first two moments of the RNPV - as a function of all possible parameter values of the model which summarizes customer behaviour (default, prepayment), making it dependent on the stochastic dynamics of market conditions. Appropriate sensitivity analyses can be performed.


The paper is structured as follows. In \autoref{sec1}, we describe how the evolution of the credit environment over time can be modeled as a Markov Chain; \autoref{sec2} introduces the model: we detail how the event of default, the recovery rate and the prepayment depend on the credit market conditions. In \autoref{sec3} we propose a procedure to calculate exactly the first two moments of the RNPV at loan level, while \autoref{sec4} generalizes such evaluation for a portfolio of risky exchangeable loans. In \autoref{sec5}, we show how the model works by a numerical application inspired by real data. We also perform a sensitivity analysis of the first two RNPV moments with respect to the evaluation rate and with respect to the portfolio dimension. \autoref{appendiceA} presents an alternative method - the \textit{backward looking} approach - for computing the RNPV's first two moments without relying on probabilities.

\section{The Evolution of the Credit Market Conditions}
\label{sec1}
We assume that the evolution of credit market conditions over time can be described by a sequence of states, each taking values from a finite set. We consider a time discrete model, and for simplicity, we assume that the state can change on a monthly basis. At time 0, which is  when the lender decides to grant a loan and the profitability of the loan must be assessed, the credit market conditions are assumed to be in one of the possible states, known at that time. At the end of the first period, just before time 1, the state may change, and this process continues for each subsequent period according to a Markov Chain.
\\
To describe how the model works and simplify the notation, we consider a set of only two states for the credit market conditions: \textit{bad} (B) and \textit{good} (G).
With two possible states for each period, a Markov Chain can be described by two sequences of parameters $(b{(1)} , b{(2)} , ...)$ and $(g{(1)}, g{(2)} , ...)$ with a proper length to cover the term of the loan.
The parameters represent the probability of persistence of the states B or G,
respectively from time $h - 1$ to time $h$, with $h = 1, 2, ...$.

Specifically, 
\begin{align}
b(h) &= \mathbb{P}(S_h = B \mid S_{h-1}=B), \\
g(h) &= \mathbb{P}(S_h = G \mid S_{h-1}=G)
\end{align}

with $S_{h-1}$ the random credit market condition at time $h-1$, that holds from the $h$-period and that may change at time $h$, the end of the $h$-period. At time 0, we suppose that $S_0$ is known with certainty and in this case $S_0=B$ or, alternatively, $S_0=G$. If the parameters $b(h)$ or $g(h)$ are close to 1 the persistence of the state is more and more likely. A similar meaning for $1-b(h)$  and $1-g(h)$. The collected lengths of the cycles related to such states allow
efficient point estimates for the parameters $b(h)$ and $g(h)$. For example, if the
average length of consecutive "bad" states is $m$, held $b(h)$ costant, we can estimate $1-b(h)=\frac{1}{m}$. For more details on how to perform and to estimate
such parameters from real datasets we refer to Quirini and Vannucci \cite{quirini2014hmm}.
\pagebreak

\section{The Model Setting}
\label{sec2}
We focus our attention at single loan level: the amount $w$ is granted at time 0
and the borrower is obliged to the contractual sequence of instalments $r_1, r_2,..., r_n$ at times $1, 2, ..., n$, with $n$ the term of the loan. We suppose that the
repayment times are on a monthly basis, in line with the time unit chosen for
the evolution of the credit market conditions.
The contractual internal rate of return, $x$, is linked to the previous cash-flow by
\begin{align}
w=\sum_{h=1}^{n} r_h(1+x)^{-h}
\end{align}
In order to put the problem under a quite general view, the certain event is partioned in $3n$ events, conditioning to the initial state $s \in \{B, G\}$. Among these events, $2n$ ones are referred to the default and its time: they are $\{A_{s,B}(h)\}_{h=1}^n$ and $\{A_{s,G}(h)\}_{h=1}^n$ if the default is observed, respectively, when the credit market conditions are in the $B$ state or in the $G$ one. There are $n - 1$ elementary events, $\{C_s(h)\}_{h=1}^{n-1}$, referred to the prepayment, while the event $E_s$ is referred to the case of a regular repayment.
In details, such events are:

\begin{itemize}
    \item \( A_{s,B}(h) \): The loan, started in state \( s \) at time 0, is in the \( B \) state during the \( h \)-th period (the interval \([h - 1, h)\)), and it defaults exactly at time \( h \), for \( h = 1, 2, \dots, n \).
    \item \( A_{s, G}(h) \): The loan, started in state \( s \) at time 0, is in the \( G \) state during the \( h \)-th period (the interval \([h - 1, h)\)), and it defaults exactly at time \( h \), for \( h = 1, 2, \dots, n \).
    \item \( C_s(h) \): The loan, started in state \( s \) at time 0, prepays at time \( h \), for \( h = 1, 2, \dots, n - 1 \).
\end{itemize}

Note that the case in which the loan prepays at maturity, $n$, would be meaningless. The considered events are $3n-1$ and are mutually incompatible.

We adjoin the event
\begin{center}
    $E_s$: "The loan is paid back regularly at maturity $n$".
\end{center}
 Such event is linked to the previous ones by the relationship 

\[
E_s = \overline{\left( \bigcup_{h=1}^{n} A_{s,B}(h) \cup \bigcup_{h=1}^{n} A_{s,G}(h) \cup \bigcup_{h=1}^{n-1} C_{s}(h) \right)}
\]
to obtain a partition of certain event into $3n$ events.
\\
If we suppose that $T$ stands for the random time when the loan defaults, prepays or concludes regularly, the possible values for $T$ are $1, 2, ..., n$. We assume that the probabilities linked to default and those linked to prepayment depend on the realizations of the Markov Chain, related to the stochastic dynamics of the credit market conditions, as follows
\begin{align}
\lambda_B(h)=\mathbb{P}(A_{s, B}(h)| T > h-1, S_{h-1}=B)  \quad h=1,2,...,n;\\
\lambda_G(h)=\mathbb{P}(A_{s, G}(h)| T > h-1, S_{h-1}=G) \quad h=1,2,..., n;\\
\mu_B(h)=\mathbb{P}(C_s(h)|T>h-1,S_{h-1}=B) \quad h=1, 2, ..., n-1;\\
\mu_G(h)=\mathbb{P}(C_s(h)|T>h-1,S_{h-1}=G) \quad  h=1, 2, ..., n-1.
\end{align}
The distinction of the event of default with respect to the state of the current credit market conditions allows to consider different levels for the recovery of the exposure according to those conditions. So, let 
\begin{center}
$Z_B(h)$ and $Z_G(h)$, $h=1,2,...,n$
\end{center}
represent the random recovery rates of exposure at default, with values in $[0,1]$, respectively, under $B$ or $G$, the state in the $i$-th period.
\\
With "exposure at default at time $h$", in symbols $\phi(h)$, we mean the present value, evaluated at the contractual internal rate $x$, of the instalments that the borrower has not paid at time $h$, including the instalment due at this time; that is
\begin{align}
\phi(h):= \sum_{j=h}^{n}r_j(1+x)^{-(j-h)}
\end{align}
For instance $\phi(1)=w(1+x)$, $\phi(n)=r_n$.

In the case of prepayment at time $h = 1, 2, ..., n-1$, we assume that the borrower pays
\begin{center}
$\gamma(h)\phi(h)$
\end{center}
with $\gamma(h) \geq 1$ and with a charge of $\gamma(h)-1$ due to prepayment: we do not distinguish between the two different market conditions in the $h$-th period when modeling prepayment, because the prepaid exposure does not depend on these conditions.

\section{The \textit{RNPV} first two moments for a single loan}
\label{sec3}
The profitability of a single loan can be assessed through its random net present value (\textit{RNPV}), evaluated at time 0 using a monthly discount rate of $y$, which is typically lower than the contractual internal rate $x$.

Let $V_B(0)$ and $V_G(0)$ represent the \textit{RNPV}s when, at time 0, the credit market starts in state B or G, respectively.

By the way, if default or prepayment risks were not present, the NPV, that the lender would gain with certainty for the loan $v_0$, would be
\begin{align}
v_0=\sum_{j=1}^{n} r_j(1+y)^{-j}-w=\sum_{j=1}^{n} r_j(1+y)^{-j}-\sum_{j=1}^{n} r_j(1+x)^{-j}=\sum_{j=1}^{n}r_j\bigg[(1+y)^{-j}-(1+x)^{-j}\bigg]
\end{align}
In our model the random variables $V_s(0)$, with $s \in \{B, G\}$ the initial state, can assume the following probability distribution:
\[
V_s(0) =
\left\{
\begin{aligned}
&\sum_{j=1}^{n} r_j(1+y)^{-j}-w, && \text{with probability } \mathbb{P}(E_s)\\
&\sum_{j=1}^{h-1} r_j(1+y)^{-j}+(1+y)^{-h}Z_B(h)\phi(h)-w, && \text{with probability } \mathbb{P}(A_{s,B}(h)), \quad h=1, \dots, n\\
&\sum_{j=1}^{h-1} r_j(1+y)^{-j}+(1+y)^{-h}Z_G(h)\phi(h)-w, && \text{with probability } \mathbb{P}(A_{s,G}(h)), \quad h=1, \dots, n\\
&\sum_{j=1}^{h-1} r_j(1+y)^{-j}+(1+y)^{-h}\gamma(h)\phi(h)-w, && \text{with probability } \mathbb{P}(C_s(h)), \quad h=1, \dots, n
\end{aligned}
\right.
\]

The expected value of $V_s(0)$ is:
\begin{align}
\mathbb{E} \left[ V_s(0) \right] = 
&\; \mathbb{P}(E_s) \left( \sum_{j=1}^{n} r_j (1+y)^{-j} - w \right) +\nonumber \\
&+ \sum_{h=1}^{n} \mathbb{P}(A_{s,B}(h)) \left( \sum_{j=1}^{h-1} r_j (1+y)^{-j} + (1+y)^{-h} \mathbb{E}[Z_B(h)] \, \phi(h) - w \right) +\nonumber \\
&+ \sum_{h=1}^{n} \mathbb{P}(A_{s,G}(h)) \left( \sum_{j=1}^{h-1} r_j (1+y)^{-j} + (1+y)^{-h} \mathbb{E}[Z_G(h)] \, \phi(h) - w \right) +\nonumber \\
& + \sum_{h=1}^{n-1} \mathbb{P}(C_s(h)) \left( \sum_{j=1}^{h-1} r_j (1+y)^{-j} + (1+y)^{-h} \gamma(h) \, \phi(h) - w \right).
\end{align}

If we set $w=0$ in the above formulas, then we will find the first two moments of the random present value, at rate $y$, of the instalment actually paid. If we indicate such present values by $R_s(0)$ with $s \in \{B, G\}$ then:
\begin{equation}
    R_s(0)=V_s(0)+w
\end{equation}
\begin{equation}
    \mathbb{E}[R_s(0)]=\mathbb{E}[V_s(0)]+w
\end{equation}
\begin{equation}
    \mathbb{V}[R_s(0)]=\mathbb{E}[V_s(0)^2]-(\mathbb{E}[V_s(0)])^2
\end{equation}
\begin{equation}
    \sqrt{\mathbb{V}[R_s(0)]}=\sqrt{\mathbb{V}[V_s(0)]}=\sqrt{\mathbb{E}[V_s(0)^2]-(\mathbb{E}[V_s(0)])^2}
\end{equation}
The comparison between $\mathbb{E}[V_s(0)]$ and the NPV, $v_0$, and the evaluation of the difference $v_0-\mathbb{E}[V_s(0)]$ allows us to estimate the expected losses due to the two competing risks depending on the initial state. Besides, the variance (and hence the standard deviation) of the variables $V_s(0)$ gives a measure of variability of such losses.

The estimation of $\mathbb{E}[V_s(0)]$ (and/or $\mathbb{V}[R_s(0)]$), for $s \in \{B,G\}$, is fundamental in order to evaluate the profitability for a loan.

For example, if $\mathbb{E}[V_B(0)]<0$ and $\mathbb{E}[V_G(0)]>0$ then the lender has an expected positive \textit{RNPV} only if the credit market at time 0 is in state $G$. For a more detailed analysis it is necessary an estimation of the variability for the \textit{RNPV}s and this can be achieved with the estimation of their variances, obtained by the second moments.

Such results, as we shall see, can be also used in evaluating profitability of a portfolio composed by many exchangeable risky loans. 

Given $R_s(0)=V_s(0)+ w$ and keeping in mind that our goal is the estimation of the first two moments for such variables, we can point out our attention on one of them, and let it be $R_s(0)$.

The formulas ... that we have just derived could give the idea that the first two moments for $R_s(0)$ could be easily computed but in such relationships we haven't described up to this point how to derive the $3n \times 2$ probabilities

\[
\mathbb{P}(E_s), \mathbb{P}(A_{s, B}(h)), \mathbb{P}(A_{s, G}(h)), \mathbb{P}(C_s(h)) \quad s \in \{B, G\}
\]

For this aim we first define
\begin{itemize}
\item $Q_{s,B}(h)$ the probability that a loan, started at time 0 with the market in state $s$, is at risk at time $h$, the beginning of the $(h+1)$-th period for $h=0, 1, ..., n-1$, with the state of the market equals $B$.
\item $Q_{s, G}(h)$ the probability that a loan, started at time 0 with the market in state $s$, is at risk at time $h$, the beginning of the $(h+1)$-th period for $h= 0, 1, ..., n-1$, with the state of market equals $G$.
\end{itemize}

For such probabilities the initial conditions satisfy
\[
Q_{BB}(0)=1 \quad \text{and} \quad Q_{BG}(0)=0,
\]
\[
Q_{GB}(0)=0 \quad\text{and}\quad Q_{GG}(0)=1
\]
and by recurisve formulas for $h=1, ..., n-1$ we have:
\[
Q_{s, B}(h)=Q_{s, B}(h-1)(1-\lambda_B(h)-\mu_B(h))b(h)+Q_{s, G}(h-1)(1-\lambda_G(h)-\mu_B(h))(1-b(h))
\]

\[
Q_{s,G}(h)=Q_{s, G}(h-1)(1-\lambda_G(h)-\mu_G(h))g(h)+Q_{s,B}(h-1)(1-\lambda_B(h)-\mu_B(h))(1-b(h))
\]

Once evaluated the $Q_{s, B}(h)$ and the $Q_{s, G}(h)$ with $h=0, 1, ..., n-1$ the probabilities of interest will be derived immediately by

\begin{equation}
\mathbb{P}(E_s)=Q_{s, B}(n-1)(1-\lambda_{B}(n))+ Q_{s, G}(n-1)(1-\lambda_{G}(n)
\end{equation}

\begin{equation}
\mathbb{P}(A_{s, B}(h))=Q_{s, B}(h-1)\lambda_{B}(h) \quad \text{for} \quad h=1, ..., n;
\end{equation}

\begin{equation}
\mathbb{P}(A_{s, G}(h))=Q_{s, G}(h-1)\lambda_{G}(h) \quad  \quad \text{for} \quad h=1, ..., n;
\end{equation}

\begin{equation}
\mathbb{P}(C_s(h))=Q_{s, B}(h-1)\mu_{B}(h)+Q_{s,G}(h-1)\mu_{G}(h) \quad\text{for}\quad h=1, ..., n-1.
\end{equation}

Another method for evaluating the first two moments of $R_s(0)$, that avoids the probabilities calculation, can be based on a \textit{backward looking} approach, presented in \autoref{appendiceA}.

\subsection{The \textit{RNPV} First Two Moments for a Portfolio of Exchangeable Loans}
In this section we generalize the previous results to a portafolio of exchangeable risky loans. Here we examine the case of $m$ loans having the same parametrization, in such a way that the \textit{RNPV} at portfolio level can be defined as (with an obvious meaning of the involved symbols) 
\begin{equation}
    \Psi_{s, m}(0)=\sum_{i=1}^{m} V_{s, i}(0)=\sum_{i=1}^{m} R_{s, i}(0)-mw
\end{equation}
with the first moment
\begin{equation}
    \mathbb{E}[\Psi_{s, m}(0)]=m\cdot\mathbb{E}[V_s(0)]
\end{equation}
It is important to assess the value for the variance of $\Psi_s(0)$. We observe that the $m$ loans share the same credit market conditions and so their RNPVs are not stocastically independent but the assumption of conditional independence and the same parametrization of loans (Exchangeability Hypotesis) allow to calculate the variance of $\Psi_{s, m}(0)$, by knowing the variance of a RNPV of a single loan and the covariance of the RNPVs for any couple of loans. It holds for the variance of $\Psi_s(0)$
\begin{equation}
    \mathbb{V}[\Psi_{s, m}(0)]=m\mathbb{V}[R_{s,1}(0)]+m(m-1)Cov[R_{s,1}(0), R_{s,2}(0)]
\end{equation}

The problem becomes the determination of the covariance of the random present value for two loans, but due to the fact that

\begin{equation}
    Cov[R_{s,1}(0), R_{s,2}(0)]=\mathbb{E}[R_{s,1}(0)R_{s, 2}(0)]-\mathbb{E}[R_{s,1}(0)]\mathbb{E}[R_{s,2}(0)]=\mathbb{E}[R_{s,1}(0)R_{s, 2}(0)]-(\mathbb{E}[R_{s,1}(0)])^2
\end{equation}
it will be sufficient to evaluate the expected value for the product of two RNPVs referred to a couple of loans, that is $\mathbb{E}[R_{s,1}(0)R_{s, 2}(0)]$.

Also this problem can be solved with a \textit{backward looking} procedure mentioned in the previous section, that is presented in \autoref{appendiceA}.

\section{An application}
\label{sec4}
We illustrate the functioning of the model through the following example, which is inspired by findings from the analysis of real-world datasets.

\textit{The loan}

$n=60$ months, $w=8500$ euros, $r_h=190$ euros with $h=1, ..., n$ and $x=0.010179$ (annual percentage rate equals $0.129224$).

\textit{The credit market conditions}

The initial state of the market $B$ or $G$ and for transition probabilities we consider the homogeneous case
\[
b(h)=0.920 \quad\text{with}\quad h=1, ..., n
\]
\[
g(h)=0.960 \quad\text{with}\quad h=1, ..., n.
\]
This choice of probabilities reflects empirical evidence indicating that credit market conditions tend to persist for extended periods—on average, 12.5 months in the Bad state and 25 months in the Good state.

\textit{The default}
\[
\lambda_B(h)=0.006  \quad\text{with}\quad h=1, ..., n;
\]
\[
\lambda_{G}(h)=0.003 \quad\text{with}\quad h=1, ..., n.
\]

With respect to the random recovery rate, we assume that is distributed according to a beta probability density
\[
\frac{\Gamma(a+b)}{\Gamma(a)\Gamma(b)}x^{a-1}(1-x)^{b-1}
\]
with $a, b \in \mathbb{R}^{+}$ and $x \in (0,1)$.
It is known that the first two moments for this distribution are, respectively

\[
\frac{a}{a+b} \quad\text{and}\quad \frac{(a+1)a}{(a+b+1)(a+b)}
\]

Here we assume that:
\begin{itemize}
\item $\mathbb{E}[Z_B(h)]=0.25$ and $Sd[Z_B(h)]=0.20$ with $h=1,...,n$ and this can be achieved with following parametrization $a=\frac{59}{64}$, $b=\frac{177}{64}$.
\item $\mathbb{E}[Z_G(h)]=0.40$ and $Sd[Z_{G}(h)]=0.20$ with $h=1, ..., n$ and this can be achieved by setting $a=2$, $b=3$.
\end{itemize}

\textit{The prepayment}

$\mu_B(h)=0.008$, $\mu_G(h)=0.010$ and $\gamma_h=1$ with $h=1, ..., n-1$.

We have computed the values of $Q_{B, B}(h)$ and $Q_{B, G}(h)$ for $h=1, ..., 59$, related to B as initial state with the given parametrization. By means of such values, we can get
\[
\mathbb{P}(E_B)=Q_{B, B}(59)(1- \lambda_{B}(60))+Q_{B, G}(59)(1-\lambda_G(60))=
\]
\[
=0.1496\cdot(1-0.006)+0.3009\cdot(1-0.003)=0.4487
\]

\[
\mathbb{P}(A_{B,B}(30))=Q_{B,B}(29)\cdot\lambda_B(30)=0.2345\cdot0.006=0.0014
\]
\[
\mathbb{P}(A_{B, G}(30))=Q_{B, G}(29)\cdot \lambda_G(30)=0.4393\cdot0.003=0.0013
\]

\[
\mathbb{P}(C_B(30))=Q_{B, B}(29)\cdot \mu_B(30)+Q_{B, G}(29)\cdot\mu_G(30)=
\]
\[
=0.2345\cdot0.008+0.4393\cdot0.010=0.0063
\]

and so on for all the others $3 \cdot 60 -4 =176$ probabilities of interest if B is the initial state.

Choosing as annual discount rate $0.040, 0.050, 0.060, 0.070, 0.08$ and as portfolio size $m=1, 8, 64, 512, 4096$ we obtain the expected values, the standard deviations for the RNPVs and their coefficients of variation. 

To this aim we have used the recursive \textit{backward looking} formulas determined in Appendix.

The results are reported in the next five tables (monetary values are approximated to integer values and the monetary unit are omitted).

\begin{table}[H]
\centering
\caption{ Annual discount rate: 0.04 \\
Contractual NPV: 1835 \\
Covariance between any couple of loans: $cov_B = 19928$, $cov_G = 15176$ \\
Linear correlation coefficients: $B = 0.00491$, $G = 0.00537$ }

\vspace{0.3cm}

\begin{tabular}{rrrrrrr}
\toprule
$m$ & $\mathbb{E}[\Psi_{B,m}(0)]$ & $Sd[\Psi_{B,m}(0)]$ & $\frac{\sigma[\Psi_{B,m}(0)]}{\mathbb{E}[\Psi_{B, m}(0)]}$ & $\mathbb{E}[\Psi_{G,m}(0)]$ & $Sd[\Psi_{G,m}(0)]$ & $\frac{\sigma[\Psi_{G,m}(0)]}{\mathbb{E}[\Psi_{G, m}(0)]}$ \\
\midrule
1     & 757    & 2014   & 2.66  & 919    & 1681   & 1.83  \\
8     & 6055   & 5793   & 0.96  & 7353   & 4843   & 0.66  \\
64    & 48439  & 18437  & 0.38  & 58826  & 15556  & 0.26 \\
512   & 387511 & 85383  & 0.22  & 470609 &  73601 & 0.16 \\
4096  & 3100088& 592340 & 0.19  & 3764868 & 515869& 0.14\\
\bottomrule
\end{tabular}
\end{table}
\vspace{0.5cm}

\vspace{0.5cm}

\begin{table}[H]
\centering
\caption{Annual discount rate 0.05\\Contractual \textit{NPV} 1595\\
Covariance between any couple of loans; $cov_B = 19120$; $cov_G = 14536$\\
Linear correlation coefficients $\rho_B = 0.00503$; $\rho_G = 0.00554$}

\vspace{0.3cm}

\begin{tabular}{rrrrrrr}
\toprule
$m$ & $\mathbb{E}[\Psi_{B,m}(0)]$ & $Sd[\Psi_{B,m}(0)]$ & $\frac{Sd[\Psi_{B,m}(0)]}{\mathbb{E}[\Psi_{B, m}(0)]}$ & $\mathbb{E}[\Psi_{G,m}(0)]$ & $Sd[\Psi_{G,m}(0)]$ & $\frac{\sigma[\Psi_{G,m}(0)]}{\mathbb{E}[\Psi_{G, m}(0)]}$\\
\midrule
1     & 582    & 1950   & 3.35  & 743    & 1620   & 2.18 \\
8     & 4659   & 5613   & 1.20  & 5942   & 4671   & 0.79 \\
64    & 37273  & 17903  & 0.48  & 47538  & 15054  &  0.32  \\
512   & 298186 & 83366  & 0.28  & 380300 & 0.32   & 71744 \\
4096  & 2385488& 579898 & 0.24  & 0.19   & 3042400& 15    \\
\bottomrule
\end{tabular}
\end{table}
\begin{table}[H]
\centering
\caption{Annual discount rate 0.04\\Contractual \textit{NPV} 1835\\
Covariance between any couple of loans; $cov_B = 19928$; $cov_G = 15176$\\
Linear correlation coefficients $\rho_B = 0.00491$; $\rho_G = 0.00537$}
\begin{tabular}{ccccccc}
\toprule
$m$ & $\mathbb{E}[\Psi_{B,m}(0)]$ & $Sd[\Psi_{B,m}(0)]$ & $\frac{Sd[\Psi_{B,m}(0)]}{\mathbb{E}[\Psi_{B, m}(0)]}$ & $\mathbb{E}[\Psi_{G,m}(0)]$ & $Sd[\Psi_{G,m}(0)]$ & $\frac{Sd[\Psi_{G,m}(0)]}{\mathbb{E}[\Psi_{G, m}(0)]}$\\
\midrule
1    & 757      & 2014   & 2.66 & 919      & 1681   & 1.83 \\
8    & 6055     & 5793   & 0.96 & 7353     & 4843   & 0.66 \\
64   & 48439    & 18437  & 0.38 & 58826    & 15556  & 0.26 \\
512  & 387511   & 85383  & 0.22 & 470609   & 73601  & 0.16 \\
4096 & 3100088  & 592340 & 0.19 & 3764868  & 515869 & 0.14 \\
\bottomrule
\end{tabular}
\end{table}

\vspace{1em}

\begin{table}[H]
\centering
\caption{Annual discount rate 0.05\\Contractual \textit{NPV} 1595\\
Covariance between any couple of loans; $cov_B = 19120$; $cov_G = 14536$\\
Linear correlation coefficients $\rho_B = 0.00503$; $\rho_G = 0.00554$}
\begin{tabular}{ccccccc}
\toprule
$m$ & $\mathbb{E}[\Psi_{B,m}(0)]$ & $Sd[\Psi_{B,m}(0)]$ & $\frac{Sd[\Psi_{B,m}(0)]}{\mathbb{E}[\Psi_{B, m}(0)]}$ & $\mathbb{E}[\Psi_{G,m}(0)]$ & $Sd[\Psi_{G,m}(0)]$ & $\frac{Sd[\Psi_{G,m}(0)]}{\mathbb{E}[\Psi_{G, m}(0)]}$\\
\midrule
1    & 582      & 1950   & 3.35 & 743      & 1620   & 2.18 \\
8    & 4659     & 5613   & 1.20 & 5942     & 4671   & 0.79 \\
64   & 37273    & 17903  & 0.48 & 47538    & 15054  & 0.32 \\
512  & 298186   & 83366  & 0.28 & 380300   & 71744  & 0.19 \\
4096 & 2385488  & 579898 & 0.24 & 3042400  & 504546 & 0.17 \\
\bottomrule
\end{tabular}
\end{table}

\begin{table}[H]
\centering
\caption{Annual discount rate 0.06\\Contractual \textit{NPV} 1366\\
Covariance between any couple of loans; $cov_B = 18424$; $cov_G = 14072$\\
Linear correlation coefficients $\rho_B = 0.00515$; $\rho_G = 0.00575$}
\begin{tabular}{ccccccc}
\toprule
$m$ & $\mathbb{E}[\Psi_{B,m}(0)]$ & $Sd[\Psi_{B,m}(0)]$ & $\frac{Sd[\Psi_{B,m}(0)]}{\mathbb{E}[\Psi_{B, m}(0)]}$ & $\mathbb{E}[\Psi_{G,m}(0)]$ & $Sd[\Psi_{G,m}(0)]$ & $\frac{Sd[\Psi_{G,m}(0)]}{\mathbb{E}[\Psi_{G, m}(0)]}$\\
\midrule
1    & 414      & 1891   & 4.56 & 573      & 1564   & 2.73 \\
8    & 3315     & 5443   & 1.64 & 4583     & 4512   & 0.98 \\
64   & 26517    & 17409  & 0.66 & 36662    & 14604  & 0.40 \\
512  & 212134   & 81551  & 0.38 & 293295   & 70242  & 0.24 \\
4096 & 1697068  & 568921 & 0.34 & 2346356  & 496034 & 0.21 \\
\bottomrule
\end{tabular}
\end{table}

\vspace{1em}

\begin{table}[H]
\centering
\caption{Annual discount rate 0.07\\Contractual \textit{NPV} 1145\\
Covariance between any couple of loans; $cov_B = 17776$; $cov_G = 13456$\\
Linear correlation coefficients $\rho_B = 0.00528$; $\rho_G = 0.00589$}
\begin{tabular}{ccccccc}
\toprule
$m$ & $\mathbb{E}[\Psi_{B,m}(0)]$ & $Sd[\Psi_{B,m}(0)]$ & $\frac{Sd[\Psi_{B,m}(0)]}{\mathbb{E}[\Psi_{B, m}(0)]}$ & $\mathbb{E}[\Psi_{G,m}(0)]$ & $Sd[\Psi_{G,m}(0)]$ & $\frac{Sd[\Psi_{G,m}(0)]}{\mathbb{E}[\Psi_{G, m}(0)]}$\\
\midrule
1    & 252      & 1835   & 7.27 & 409      & 1511   & 3.70 \\
8    & 2019     & 5285   & 2.62 & 3272     & 4362   & 1.33 \\
64   & 16149    & 16945  & 1.05 & 26180    & 14159  & 0.54 \\
512  & 129193   & 79841  & 0.62 & 209437   & 68485  & 0.33 \\
4096 & 1033540  & 558524 & 0.54 & 1675496  & 484827 & 0.29 \\
\bottomrule
\end{tabular}
\end{table}

\begin{table}[H]
\centering
\caption{Annual discount rate 0.08\\Contractual \textit{NPV} 933\\
Covariance between any couple of loans; $cov_B = 17184$; $cov_G = 12968$\\
Linear correlation coefficients $\rho_B = 0.00541$; $\rho_G = 0.00606$}
\begin{tabular}{ccccccc}
\toprule
$m$ & $\mathbb{E}[\Psi_{B,m}(0)]$ & $Sd[\Psi_{B,m}(0)]$ & $\frac{Sd[\Psi_{B,m}(0)]}{\mathbb{E}[\Psi_{B, m}(0)]}$ & $\mathbb{E}[\Psi_{G,m}(0)]$ & $Sd[\Psi_{G,m}(0)]$ & $\frac{Sd[\Psi_{G,m}(0)]}{\mathbb{E}[\Psi_{G, m}(0)]}$\\
\midrule
1    & 96      & 1782   & 18.55 & 251      & 1463   & 5.83 \\
8    & 769     & 5136   & 6.68  & 2009     & 4224   & 2.10 \\
64   & 6151    & 16512  & 2.68  & 16070    & 13757  & 0.86 \\
512  & 49207   & 78247  & 1.59  & 128561   & 66996  & 0.52 \\
4096 & 393656  & 548857 & 1.39  & 1028488  & 475688 & 0.46 \\
\bottomrule
\end{tabular}
\end{table}

These results can be readily interpreted in light of their economic and financial implications. An increase in the annual discount rate leads to a decline in all risk indicators, indicating reduced profitability across the evaluated portfolios. In practical settings, such findings may assist lenders in identifying boundaries in the 
$(y,m)$-plane that delineate regions of significantly different levels of profitability.

We also note that the increase of the portfolio size (in the above tables from  the single transaction to a portfolio having 4096 loans with 8 as multiplier) reduces the related coefficient of variation
\[
\frac{Sd[\Psi_{s, m}(0)]}{\mathbb{E}[\Psi_{s, m}(0)]} \quad s \in \{B, G\}
\]
but the positive dependence among the RNPVs, measured by small but not null linear correlation coefficients, doesn't allow to reduce aribitraily its value.
The coefficients of variation are decreasing as the proftfolio's size increases but they are larger than their limit value.
\[
\lim_{m\rightarrow\infty} \frac{Sd[\Psi_{s, m}(0)]}{\mathbb{E}[\Psi_{s, m}(0)]} =\frac{\sqrt{cov(R_{s,1}(0), R_{s.2}(0))}}{\mathbb{E}[\Psi_{s, m}(0)]}
\]

This limit value can be used to estimate the systematic risk arising from the fact that all the loans share a common economic-financial environment. Next table reports the limit value for the coefficient of variation (with three decimal place) by modifying the annual discount rate and the initial credit market condition.

\begin{table}[ht]
\centering
\begin{tabular}{|c|c|c|}
\hline
\textbf{Annual discount rate} & 
$\dfrac{\sqrt{\operatorname{cov}(R_{B,1}(0), R_{B,2}(0))}}{\mathbb{E}[V_B(0)]}$ & 
$\dfrac{\sqrt{\operatorname{cov}(R_{G,1}(0), R_{G,2}(0))}}{\mathbb{E}[V_G(0)]}$ \\
\hline
0.04 & 0.186 & 0.134 \\
0.05 & 0.238 & 0.162 \\
0.06 & 0.328 & 0.207 \\
0.07 & 0.529 & 0.284 \\
0.08 & 1.389 & 0.462 \\
\hline
\end{tabular}
\end{table}

Such values are close to the ones obtained previously with $m=4096$. This fact suggests that this portfolio's size is sufficient to allow the maximum possible reduction of the specific risk.

\section{Conclusions}
\label{sec5}

In this paper, we present a model to measure the profitability of a portfolio of exchangeable loans, considering the two most significant financial risks: default and prepayment. These risks are evaluated in relation to the dynamic conditions of the credit market. We have demonstrated, particularly through recursive formulas, how to compute the first two moments of the RNPV for both individual loans and portfolios. A practical representative case is also discussed, with numeric evidence illustrating the results. This allows some useful sensitivity analysis with respect to the evolution rate and the size of the portfolio.

\appendix
\section{\textit{Backward looking approach}}
\label{appendiceA}
We present an alternative method for evaluating the first two moments of $R_s(0)$, avoiding probabilities. This approach is called \textit{backward looking}.

We first show how the method can directly yield the exact values of $\mathbb{E}[R_B^k(0)]$ and $\mathbb{E}[R_G^k(0)]$ for $k = 1, 2$. We outline the steps of the procedure below.
\begin{enumerate}
    \item Suppose we have the following two sequences of random variables with $h=1, ..., n$:
    \begin{itemize}
        \item[i.] $R_B(h-1)$ the present value of the instalments $r_h ,..., r_n$ evaluated at $h-1$, knowing that at time $h-1$ the credit market condition is B.
        \item[ii.] $R_G(h-1)$ the present value of the instalments $r_h ,..., r_n$ evaluated at $h-1$, knowing that at time $h-1$ the credit market condition is G.
    \end{itemize}
    Note that for $h=1$ we obtain the two random variables of interest: $R_B(0)$ and $R_G(0)$.
    \item Given the periodic discount factor $v=(1+y)^{-1}$, it is possible to compute the initial values for the \textit{backward looking} procedure related to time $n-1$.
    \begin{align}
    \mathbb{E}[R_B(n-1)]=v\cdot\big[\lambda_B(n)\cdot r_n \cdot \mathbb{E}[Z_B(n)]+ \big(1-\lambda_B(n)\big)\cdot r_n\big]
    \tag{23}
    \\
    \mathbb{E}[R_G(n-1)]=v\cdot\big[\lambda_G(n)\cdot r_n \cdot \mathbb{E}[Z_G(n)]+ \big(1-\lambda_G(n)\big)\cdot r_n\big] \tag{24}
    \\
    \mathbb{E}[R_B^2(n-1)]=v^2\cdot\big[\lambda_B(n)\cdot r_n^2 \cdot \mathbb{E}[Z^2_B(n)]+ \big(1-\lambda_B(n)\big)\cdot r^2_n\big] \tag{25}
    \\
    \mathbb{E}[R_G^2(n-1)]=v^2\cdot\big[\lambda_G(n)\cdot r_n^2 \cdot \mathbb{E}[Z^2_G(n)]+ \big(1-\lambda_G(n)\big)\cdot r^2_n\big] \tag{26}
    \end{align}
    \item At this point, we may consider at time periods $n-2, ..., h, h-1, ..., 0$ in order to find out the values of 
    \[
    \mathbb{E}[R_B(n-1)],  \mathbb{E}[R_G(n-1)], \mathbb{E}[R_B^2(n-1)], \mathbb{E}[R_G^2(n-1)] 
    \]
    once we know the values of 
    \[
    \mathbb{E}[R_B(h)],  \mathbb{E}[R_G(h)], \mathbb{E}[R_B^2(h)], \mathbb{E}[R_G^2(h)] 
    \]
    In detail we obtain:
    \begin{align}
    \mathbb{E}\left[ R_B(h-1) \right] &= v \cdot \Bigg( 
    \big( \lambda_B(h)\,\phi(h)\,\mathbb{E}\left[Z_B(h)\right] + \mu_B(h)\,\gamma(h)\,\phi(h)\big) \notag \\
    &\quad+ \big(  (1 - \lambda_B(h) - \mu_B(h)\big)\big(r_h + b(h)\,\mathbb{E}\left[R_B(h)\right] + (1 - b(h))\,\mathbb{E}\left[R_G(h)\right] \big) \Bigg), \tag{27}
\\
\mathbb{E}\left[ R_G(h-1) \right] &= v \cdot \Bigg(\big( \lambda_G(h)\,\phi(h)\,\mathbb{E}\left[Z_G(h)\right] + \mu_B(h)\,\gamma(h)\,\phi(h)\big) \notag \\
&\quad+ \big(  (1 - \lambda_G(h) - \mu_G(h)\big)\big(r_h + g(h)\,\mathbb{E}\left[R_B(h)\right] + (1 - g(h))\,\mathbb{E}\left[R_G(h)\right] \big) \Bigg), \tag{28}
\\
\mathbb{E}\left[ R_B^2(h-1) \right] &= v^2 \cdot \Bigg[ 
\left( \lambda_B(h)\,\phi^2(h)\,\mathbb{E}\left[Z_B^2(h)\right] + \mu_B(h)\,\gamma^2(h)\,\phi^2(h) + \right. \notag \\
&\quad + \left. (1 - \lambda_B(h) - \mu_B(h)) \cdot 
\Bigg( r_h^2 + b(h)\left(2r_h \mathbb{E}\left[R_B(h)\right] +\mathbb{E}\left[R_B^2(h)\right] \right) \right. \Bigg) \notag \\
&\quad \left. \left. + (1 - g(h))\left(2r_h \mathbb{E}\left[R_G(h)\right] + \mathbb{E}\left[R_G^2(h)\right] \right) \right.
\right) \Bigg], \tag{29}
\\
\mathbb{E}\left[ R_G^2(h-1) \right] &= v^2 \cdot \Bigg[ 
\left( \lambda_G(h)\,\phi^2(h)\,\mathbb{E}\left[Z_G^2(h)\right] + \mu_G(h)\,\gamma^2(h)\,\phi^2(h) + \right. \notag \\
&\quad + \left. (1 - \lambda_G(h) - \mu_G(h)) \cdot 
\Bigg( r_h^2 + g(h)\left(2r_h \mathbb{E}\left[R_G(h)\right] +\mathbb{E}\left[R_G^2(h)\right] \right) \right. \Bigg) \notag \\
&\quad \left. \left. + (1 - g(h))\left(2r_h \mathbb{E}\left[R_B(h)\right] + \mathbb{E}\left[R_B^2(h)\right] \right) \right.
\right) \Bigg]. \tag{30}
\end{align}
For the standard deviation for $R_s(0)$ or $V_s(0)$ with $s \in \{B, G\}$ we get 
\[
V[R_s(0)]=\sqrt{\mathbb{E}[R^2_s(0)]-\mathbb{E}^2[R_s(0)]}
\]
\end{enumerate}
This procedure can be applied in order to evaluate the expected value for the product of two RNPVs referred to a couple of loans, considering
\[
\mathbb{E}[R_{B,1}(h)R_{B, 2}(h)] \quad\text{and}\quad \mathbb{E}[R_{G,1}(h)R_{G, 2}(h)] \text{for} h=n-1, n-2, ..., 1, 0 
\]
Note that the \textit{probabilities approach} would require a finer partition of the certain event into $(3n)^2$ elementary events!

Under the \textit{backward looking} approach we evaulate the initial values related to time $n-1$, using the Exchangeability Hypotesis.

\begin{align}
\mathbb{E}[R_{B,1}(n-1)R_{B, 2}(n-1)]= (v\varphi(n))^2
\cdot\bigg(\lambda_B(n)\mathbb{E}[Z_{B,1}(n)] + \big( 1- \lambda_B(n)\big) \bigg)^2
\\
\mathbb{E}[R_{G,1}(n-1)R_{G, 2}(n-1)]= (v\varphi(n))^2
\cdot\bigg(\lambda_G(n)\mathbb{E}[Z_{G,1}(n)] + \big( 1- \lambda_G(n)\big) \bigg)^2
\end{align}
Afterward for the times $n-2, ..., h, h-1, ...,0$ we evaluate $\mathbb{E}[R_{B, 1}(h-1)R_{B, 2}(h-1)]$, $\mathbb{E}[R_{G, 1}(h-1)R_{G, 2}(h-1)]$ once $\mathbb{E}[R_{B, 1}(h)R_{B, 2}(h)]$, $\mathbb{E}[R_{G, 1}(h-1)R_{G, 2}(h-1)]$, $\mathbb{E}[R_{B,1 }(h)]$ and $\mathbb{E}[R_{G,1}(h)]$ have been calculated. Noting the use Exchangeability Hypotesis, we find out:
\begin{align}
\mathbb{E}[R_{B, 1}(h-1)R_{B, 2}(h-1)] = v^2\varphi^2(h)\bigg(\lambda_B(h)\mathbb{E}[Z_{B,1}(h)]+\mu_B(h)\gamma(h)\bigg)^2 +
2\varphi(h)\big(1-\lambda_B(h)-\mu_B(h)\big)\cdot \notag\\
 \cdot\bigg(r_h + b(h)\mathbb{E}[R_{B,1}(h)]+(1-b(h))\mathbb{E}[R_{G,1}(h)]\bigg)\bigg(\mathbb{E}[Z_{B,1}(h)]\lambda_B(h)+\gamma(h)+\mu_B(h)\bigg)+\notag\\
 +(1-\lambda_B(h)-\mu_B(h))^2\cdot\bigg(
 r_h^2+b(h)\big(2\cdot r_h \mathbb{E}[R_{B,1}(h)]+\mathbb{E}[R_{B,1}(h)R_{B,2}(h)]\big)+\notag\\
 +(1-b(h))\big(2r_h\mathbb{E}[R_{G,1}(h)]+\mathbb{E}[R_{G,1}(h)R_{G,2}(h)]\big)\bigg)\tag{25}
 \\
 \mathbb{E}[R_{G, 1}(h-1)R_{G, 2}(h-1)] = v^2\varphi^2(h)\bigg(\lambda_B(h)\mathbb{E}[Z_{G,1}(h)]+\mu_G(h)\gamma(h)\bigg)^2 +
2\varphi(h)\big(1-\lambda_G(h)-\mu_G(h)\big)\cdot \notag\\
 \cdot\bigg(r_h + g(h)\mathbb{E}[R_{G,1}(h)]+(1-g(h))\mathbb{E}[R_{B,1}(h)]\bigg)\bigg(\mathbb{E}[Z_{G,1}(h)]\lambda_G(h)+\gamma(h)+\mu_G(h)\bigg)+\notag\\
 +(1-\lambda_G(h)-\mu_B(h))^2\cdot\bigg(
 r_h^2+b(h)\big(2\cdot r_h \mathbb{E}[R_{G,1}(h)]+\mathbb{E}[R_{G,1}(h)R_{G,2}(h)]\big)+\notag\\
 +(1-g(h))\big(2r_h\mathbb{E}[R_{B,1}(h)]+\mathbb{E}[R_{B,1}(h)R_{B,2}(h)]\big)\bigg)\tag{26}
\end{align}
The standard deviation of $\Psi_s(0)$ for $s\in \{B, G\}$ is 
\begin{align}
\sqrt{mV[R_{s,1}(0)]+m(m-1)\cdot(\mathbb{E}[R_{s,1}(0)R_{s,2}(0)]-\mathbb{E}^2[R_{s,1}(0)])}
\end{align}
\newpage
\bibliographystyle{unsrt} 
\bibliography{references}

\begin{thebibliography}{10}

\bibitem{quirini2011monitoring}
L.~Quirini and L.~Vannucci.
\newblock Monitoring creditworthiness: Markov chains and replication portfolios, 2011.
\newblock Presentation, Credit Scoring and Credit Control XII, Business School University of Edinburgh.

\bibitem{crowder2001}
M.~Crowder.
\newblock {\em Classical Competing Risks}.
\newblock Chapman \& Hall, 2001.

\bibitem{crowder2012}
M.~Crowder.
\newblock {\em Multivariate Survival Analysis and Competing Risks}.
\newblock Chapman \& Hall, 2012.

\bibitem{BSA2012}
J.~Bayersman, M.~Schumacher, and A.~Allignol.
\newblock {\em Competing Risks and Multistate Models with R}.
\newblock Springer, 2012.

\bibitem{narain1992survival}
B.~Narain.
\newblock Survival analysis and the credit granting decision.
\newblock In {\em Credit Scoring and Credit Control}. Oxford University Press, 1992.

\bibitem{stepanova2002survival}
M.~Stepanova and L.~Thomas.
\newblock Survival analysis methods for personal loan data.
\newblock {\em Operational Research}, 50(2):277--289, 2002.

\bibitem{bellotti2009credit}
T.~Bellotti and J.~Crook.
\newblock Credit scoring with macroeconomic variables using survival analysis.
\newblock {\em Journal of the Operational Research Society}, 60:1699--1707, 2009.

\bibitem{cyert1962estimation}
R.~Cyert, J.~Davidson, and G.~Thompson.
\newblock Estimation of the allowance for doubtful accounts by markov chains.
\newblock {\em Management Science}, 8:319, 1962.

\bibitem{crowder2005}
M.~Crowder, M.~Davis, and G.~Giampieri.
\newblock Analysis of default data using hidden markov models.
\newblock {\em Quantitative Finance}, 5(1):273--274, 2005.

\bibitem{zucchini2009hidden}
W.~Zucchini and I.~L. MacDonald.
\newblock {\em Hidden Markov Models for Time Series}.
\newblock Chapman \& Hall, 2009.
\newblock Op.[].

\bibitem{quirini2014hmm}
L.~Quirini and L.~Vanucci.
\newblock Creditworthiness dynamics and hidden markov models.
\newblock {\em Journal of the Operational Research Society}, 65(3), March 2014.

\bibitem{firdoos}
Noor Firdoos.
\newblock Using hidden markov model to monitor possible loan defaults in banks.
\newblock {\em International Journal of Economics and Business Administration}, 2020.

\bibitem{yu2018hmmcredit}
Feng-Hui Yu, Jiejun Lu, Jia-Wen Gu, and Wai-Ki Ching.
\newblock Modeling credit risk with hidden markov default intensity.
\newblock {\em Computational Economics}, 2018.
\newblock Accepted: 29 October 2018 / Published online: 20 November 2018.

\bibitem{bluemke2022shmm}
Oliver Blümke.
\newblock A structural hidden markov model for forecasting scenario probabilities for portfolio loan loss provisions.
\newblock {\em Knowledge-Based Systems}, 249:108934, August 2022.

\end{thebibliography}

\end{document}